\PassOptionsToPackage{table,xcdraw,dvipsnames}{xcolor}
\documentclass[sigconf]{acmart}

\AtBeginDocument{%
  }

\usepackage{amsmath,amsfonts}
\usepackage{algorithmic}
\usepackage{graphicx}
\usepackage{textcomp}
\usepackage{float}
\usepackage{xspace}
\usepackage{adjustbox}
\usepackage{booktabs}
\usepackage{makecell}
\usepackage{multirow}
\usepackage{svg}
\usepackage{fontawesome5}\usepackage{tabularx}
\usepackage{array}
\usepackage{hyperref}
\usepackage{caption}
\usepackage{subcaption}
\usepackage{siunitx}
\usepackage{comment}
\usepackage{tikz}
\newcommand{\numbercircle}[2][black]{%
  \tikz[baseline=(char.base)]{
    \node[draw, circle, fill=#1, text=white, inner sep=1.5pt] (char) {#2};
  }%
}

\definecolor{beige}{RGB}{245, 245, 220}
\definecolor{lightgreen}{RGB}{220, 255, 220}
\definecolor{lightred}{RGB}{255, 220, 220}

\hypersetup{
    colorlinks=true,
    linkcolor=blue,
    citecolor=blue,
    filecolor=magenta,
    urlcolor=blue
}

\newcommand{\controlflow}{\textsc{control flow}\xspace}
\newcommand{\dataflow}{\textsc{data flow}\xspace}

\newcommand{\identifier}{\textsc{identifier naming}\xspace}

\newcommand{\rqOne}{RQ$_1$: To what extent do semantics-preserving perturbations (SPP) impact the consistency of state-of-the-art ACR tools?}
\newcommand{\rqOneShort}{RQ$_1$: To what extent do SPPs impact the consistency of state-of-the-art ACR tools?}
\newcommand{\rqTwo}{RQ$_2$: Do SPP features relate to inconsistency in ACR tools?}
\newcommand{\rqThree}{RQ$_3$: Do changes in input representation reduce the inconsistency of ACR tools against SPPs?}

\copyrightyear{2026}
\acmYear{2026}
\setcopyright{cc}
\setcctype{by}
\acmConference[MSR '26]{23rd International Conference on Mining Software Repositories}{April 13--14, 2026}{Rio de Janeiro, Brazil}
\acmBooktitle{23rd International Conference on Mining Software Repositories (MSR '26), April 13--14, 2026, Rio de Janeiro, Brazil}
\acmDOI{10.1145/3793302.3793363}
\acmISBN{979-8-4007-2474-9/2026/04}

\begin{document}

\title{Consistent or Sensitive? Automated Code Revision Tools Against Semantics-Preserving Perturbations}

\author{Shirin Pirouzkhah}
\affiliation{%
	\institution{University of Zurich}
	\city{Zurich}
	\country{Switzerland}
}
\email{shirin@ifi.uzh.ch}

\author{Souhaila Serbout}
\affiliation{%
	\institution{University of Zurich}
	\city{Zurich}
	\country{Switzerland}
}
\email{souhaila.serbout@uzh.ch}

\author{Alberto Bacchelli}
\affiliation{%
	\institution{University of Zurich}
	\city{Zurich}
	\country{Switzerland}
}
\email{bacchelli@ifi.uzh.ch}

\renewcommand{\shortauthors}{Pirouzkhah et al.}

\begin{abstract}
Automated Code Revision (ACR) tools aim to reduce manual effort by automatically generating code revisions based on reviewer feedback. While ACR tools have shown promising performance on historical data, their real-world utility depends on their ability to handle similar code variants expressing the same issue---a property we define as \emph{consistency}. However, the probabilistic nature of ACR tools often compromises consistency, which may lead to divergent revisions even for semantically equivalent code variants. 

In this paper, we investigate the extent to which ACR tools maintain consistency when presented with semantically equivalent code variants. 
To do so, we first designed nine types of semantics-preserving perturbations (SPP) and applied them to \num{2032} Java methods from real-world GitHub projects, generating over \num{10}K perturbed variants for evaluation.
Then we used these perturbations to evaluate the consistency of five state-of-the-art transformer-based ACR tools. 
We found that the ACR tools' ability to generate correct revisions can drop by up to 45.3\%, when presented with semantically equivalent code. 
The closer the perturbation is to this targeted region, the more likely an ACR tool is to fail to generate the correct revision. 
We explored potential mitigation strategies that modify the input representation, but found that these attention-guiding heuristics yielded only marginal improvements, thus leaving the solution to this problem as an open research question.
\end{abstract}



\maketitle

\section{Introduction}

Code review is a software engineering process in which developers, referred to as reviewers, evaluate code changes proposed by a peer, referred to as the author, with the goal of ensuring code quality~\cite{moralesquality,bavotaquality,improvequality} and minimizing defects~\cite{bacchelliexpectations,rigbydefect,reducedefects}. This process is known to be labor-intensive and time-consuming~\cite{sadowski2018modern}. To alleviate this effort, researchers have investigated whether and how sequence-to-sequence (seq2seq) learning-based models can automate stages of the code review process, demonstrating promising performance across a variety of tasks~\cite{tufano2021towards,tufano2022,microsoft,vincentatscale,Howfar,Patanamon}. \emph{Code revision} (i.e., the act of modifying the code according to reviewer feedback) is one of these stages, and it requires understanding reviewers' natural-language feedback, identifying the relevant code fragments, and implementing functionally correct changes that address the mentioned issue. 

\begin{figure}[ht]
\centering
\includegraphics[width=1\linewidth]{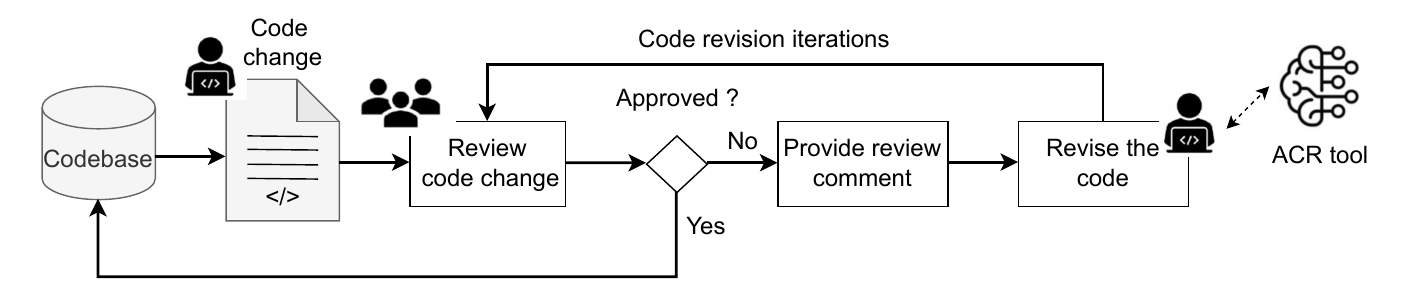}
\vspace{-0.7cm}
\caption{
 Code revision iterations and ACR tool in the process}
\label{fig:coderevision_process}
\end{figure}

In this study, we focus on the tools proposed for Automated Code Revision (ACR). An ACR tool is typically implemented as a seq2seq model that generates revised code based on the initial code and the associated review comment. ACR tools have been developed using various neural architectures, including long short-term memory networks (LSTM RNNs)~\cite{hochreiter1997long} and transformer-based architectures~\cite{vaswani2017attention}. Our study focuses on transformer-based ACR tools, as they showed the most promise for this task~\cite{tufano2022, microsoft, distilledT, lamaReviewer, nashaattowards, tufano2021towards, coditt5, pornprasit2024fine}. Compared to earlier RNN-based approaches, transformer-based models more effectively capture long-range dependencies and global context in source code and review comments through self-attention.
\autoref{fig:coderevision_process} illustrates how an ACR tool integrates into the code review workflow.

\begin{figure*}[ht]
    \centering
    \includegraphics[width=\textwidth]{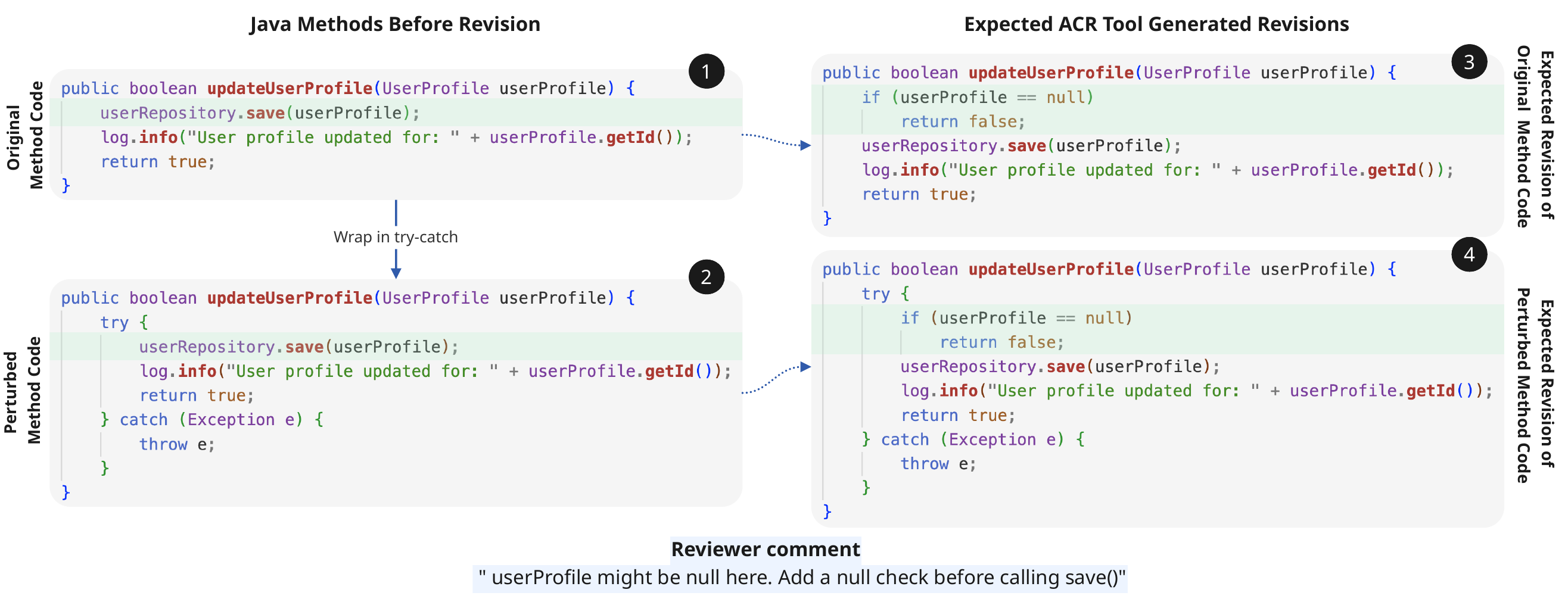}
    \caption{
    Example of original code (\numbercircle{1}), its semantically equivalent variant (\numbercircle{2}) and their expected revisions (\numbercircle{3} \& \numbercircle{4}).}
    \label{fig:consistency_matters}
\end{figure*}

ACR tools have been often evaluated in terms of accuracy, using metrics such as Exact Match (EXM), BLEU or CodeBLEU, and Levenshtein distance to measure the similarity between the generated and expected correct revision~\cite{Howfar,tufano2021towards,tufano2022}.
Beyond accuracy, ACR tools must also demonstrate \emph{consistency}: the ability to generate semantically equivalent revisions for semantically equivalent inputs~\cite{CounterfactualsForCode,ICSEcounterfactual}.
In fact, to be practically useful and to gain developers’ trust, ACR tools should ideally generate consistent revisions in response to identical reviewer feedback for the same issue, even when the underlying code has undergone minor, semantics-preserving changes (for example, renaming variables or reordering conditional statements).
\autoref{fig:consistency_matters} illustrates why consistency is an expected property of ACR tools. When an ACR tool can correctly revise the original code based on a reviewer's comment, a semantically equivalent variant of the original code (such as wrapping the method body in a \texttt{try–catch} block) should not alter the revision. In this example, wrapping a method body in a \texttt{try–catch} block keeps the semantics of the code unchanged as the catch clause re-throws the caught exception. The review comment still targets the same issue; the underlying program behavior remains unchanged, and the expected revision should remain identical. In this paper, we restrict this expectation to cases where the variant is semantics-preserving and does not address the issue described in the review comment; thus, the intended fix remains unchanged despite the structural variation.


However, consistency may be problematic for ACR tools, due to the probabilistic nature of the transformer models on which they build. In fact, prior studies have shown that large language models (LLMs), i.e., the transformer models used in ACR tools, including those applied to code, exhibit sensitivity to input variations in specific programming tasks such as code completion~\cite{CounterfactualsForCode, wang2022recode} and vulnerability detection~\cite{zhang2024attacks}. For example, LLMs can exhibit substantial sensitivity to changes such as renaming functions and variables or restructuring code blocks, even when these modifications preserve program semantics. These findings suggest that current models may only rely on surface-level syntax patterns rather than representations of semantic relationships. Yet, unlike code completion, ACR tools are expected to generate revisions that respond to natural language review comments. This makes the code revision task different than others, as ACR tools get the guidance of the reviewer's feedback.

In this paper, we evaluate the \textit{consistency} of ACR tools by measuring their sensitivity against semantics-preserving perturbations (SPP)---controlled code modifications that alter the structure or syntax of the code while keeping its behavior, input–output relations, and observable effects unchanged. We devised nine types of SPP designed to target three key programming concepts: \emph{control flow}, \emph{data flow}, and \emph{identifier naming}. To this end, we implement a framework that automatically parses Java methods into Abstract Syntax Trees (ASTs), applies one SPP at a time (to isolate the effect of each SPP), and reconstructs compilable code for evaluation. Applying SPPs cumulatively would make it difficult to determine which perturbation is associated with the model's performance in terms of consistency. To ensure diversity among ACR tools, we build on the systematic literature review by~\citet{SLRCR}, which identifies state of the art approaches to automated code revision. From this review, we select tools implemented on transformer-based backbones, covering model sizes, fine-tuning strategies, and training objectives. The final set of evaluated tools includes fine-tuned T5-small~\cite{raffel2020exploring}, LoRA-tuned LLaMA-7B~\cite{touvron2023llama,hu2022lora}, LLaMA 3.3-70B~\cite{dubey2024llama}, ChatGPT-3.5 Turbo~\cite{openai2023gpt35}, and DeepSeek V3~\cite{deepseekai2024deepseekv3}.

We use a dataset of \num{2032} real-world GitHub review instances, where each instance represents a case in which at least one ACR tool successfully generated the correct revision for the original, unperturbed code, matching exactly the change done by the original developers in response to that comment (also known as, \emph{Exact Match}). We applied nine types of SPP to every instance in this dataset, creating over \num{10}K perturbed but semantically equivalent variants for evaluation. For each model, to analyze consistency degradation across perturbation types, we selected a model-specific subset of these \num{2032} instances for which that particular model successfully produced correct revisions on the original inputs. The corresponding perturbed variants of these instances were then used to assess that model’s consistency. This approach ensured that each model's baseline performance was grounded in its own ability to correctly revise the original inputs, before testing its consistency under SPP. Since each model differs in its baseline capability, these subsets varied in size. Finally, to enable fair cross-model comparison, we constructed an intersection subset of \num{464} review instances for which all evaluated models revised the original code exactly as the original developers.
With our work, we seek to answer these research questions: 

\noindent\textbf{\rqOne}

  \noindent Different types of perturbations may pose varying degrees of inconsistency for ACR tools. We hypothesize that more complex code perturbations negatively impact the consistency with which these tools generate revisions, due to the tendency of large language models to rely on pattern repetition~\cite{bender2021dangers}, a consequence of their learning objectives and limited semantic understanding~\cite{edman2024cute}.

\noindent\textbf{\rqTwo}

\noindent Each perturbation can be described by quantifiable features such as its position within the method or its proximity to the code lines that require revision. We use these interpretable features as independent variables in a multilevel mixed-effects logistic regression model to examine whether and how they relate to the likelihood of an ACR tool failing to generate the correct revision.

\noindent\textbf{\rqThree}

\noindent We explore whether input representations can guide model attention toward the code lines that require revision and improve the consistency of ACR tools when faced with semantically equivalent code variants. We examine three approaches: (1) reinforcing attention by repeating the code lines referenced in the review comment within the review comment, (2) embedding the review comment directly as inline comments to align it with the corresponding code lines, and (3) Chain-of-Thought reasoning in prompt-based models to encourage stepwise explanation before code generation by appending a ``step-by-step explanation'' phrase to the input prompt,. These strategies test whether strengthening the connection between review feedback and the referenced code lines (by review feedback) can mitigate inconsistency or inadvertently amplify it.

\smallskip Overall, we find that ACR tools are sensitive to SPPs in the input code, with their consistency degrading by up to 45\% in the worst cases. This degradation, however, is not uniform across all perturbation types. The largest drops occur for perturbations that fundamentally restructure the control flow—rather than merely inserting dead or inactive code—such as those that reorder or reform existing logic blocks. Similarly, perturbations that alter data flow, for example by changing how variables are defined or propagated, also lead to substantial declines in consistency. The decline in consistency becomes particularly severe when perturbations are introduced near the specific lines of code referenced in the reviewer’s feedback, where models most often fail to generate the correct revision. Our attempts to mitigate this issue through attention-guiding heuristics produced only marginal improvements.

\section{Related Work}

\noindent\textit{{\textbf{Automated Code Revision (ACR) Tools. }}} 
A systematic literature review by \citet{SLRCR} categorizes the landscape of automation in code review, including tasks such as assessing review quality, classifying code changes, and, most relevant to this work, \textit{revised code generation}. This task, which they define as ``generating a revised version of a given piece of code by implementing a specific change requested by the reviewer in a natural language comment'', corresponds to the definition of ACR used in this paper's study. Their review identifies 11 key studies falling under this definition. \autoref{tab:acr_summary}
provides a summary of the 11 studies of automated code revision generation based on review comments, highlighting the models used, their exact match scores (EXM\%) if reported, BLEU scores if reported, their input format and availability of their tools. For this study, we select ACR tools that meet the following criteria: (i)~They are powered by transformer-based language models; (ii) They accept method-level code as input, rather than change diffs or other representations because our framework applies perturbations to code using JavaParser, which requires valid Java method declarations as input to ensure accurate parsing; (iii) their model checkpoints are publicly available, to ensure reproducibility and consistent evaluation.

\begin{table}[t]
\caption{Summary of LLM-based ACR studies}  
\centering
\begin{tabular}{lp{7.5cm}}
Input types & M: method code, AM: abstracted method, \\
 &  D: diff, C: review comment \\
Tool availability & \faCheckCircle~Available \quad \faTimesCircle~Not Available \quad 
\end{tabular}
\vspace{0.2cm}

\label{tab:acr_summary} 
\begin{adjustbox}{max width=\columnwidth}
\begin{tabular}{rlrrl}
\toprule
\textbf{} & \textbf{ACR tool} & \textbf{EXM\%} & \textbf{BLEU} & \textbf{Input} \\
\midrule
\faCheckCircle &  T5 small~\cite{tufano2022} & 18.88 & 80.00 & M+C \\
\faCheckCircle &  T5 (CodeT5 params)~\cite{microsoft} & 30.32 & 82.61 & D+C \\
\faCheckCircle &  Distilled CodeT5~\cite{distilledT} & -- & 85.49 & D+C \\
\faCheckCircle &  LLaMA, LoRA-tuned~\cite{lamaReviewer} & -- & 78.23 & M+C \\
\faCheckCircle & LLaMA, Prefix-tuned~\cite{lamaReviewer} & -- & 77.04 & M+C \\
\faTimesCircle &  CodeLlama, LoRA-tuned~\cite{nashaattowards} & 43.71 & 86.17 & M+C \\
\faCheckCircle &  OpenNMT\cite{tufano2021towards} & 30.72 & 95.00 & AM+C \\
\faCheckCircle &  LSTM encoder-decoder~\cite{review4repair} & 31.51 & -- & M+C \\
\faCheckCircle &  Pretrained CoditT5~\cite{coditt5}  & -- & -- & -- \\
\faTimesCircle &  Fine-tuned CoditT5~\cite{coditt5} & 43.42 & 83.92 & M+C \\
\faTimesCircle &  Fine-tuned GPT-3.5~\cite{pornprasit2024fine} & 22.16 & 82.99 & M+C \\
\faCheckCircle &  Fine-tuned Magicoder~\cite{pornprasit2024fine} & 11.14 & 69.77 & M+C \\
\faTimesCircle &  CodeT5-Base ENREFINER~\cite{lu2023improving} & 36.32 & -- & M+C \\
\faTimesCircle &  T5X framework\cite{frommgen2024resolving} & -- & -- & M+C \\
\bottomrule
\end{tabular}
\end{adjustbox}
\end{table}

\noindent\textit{\textbf{Sensitivity of LLMs to input variations.} }Previous research has shown that, in specific coding-related tasks, the output of LLMs can be inconsistent when slightly perturbed. This phenomenon highlights the sensitivity of LLMs to minimal changes, raising concerns about thereliability of their responses for other types of software engineering tasks. To systematically study these behaviors, researchers have used counterfactual mutations as a specific type of perturbations. These mutations help to uncover which aspect of the input modifications significantly influences the model's response. \citeauthor{CounterfactualsForCode} \cite{CounterfactualsForCode} introduce the Counterfactual Analysis framework to evaluate the understanding of programming concepts by LLMs in the code completion task. By generating counterfactual versions of code that only modify specific programming concepts, \citeauthor{CounterfactualsForCode} \cite{CounterfactualsForCode} revealed that LLM's grasp of programming principles is often superficial. \citeauthor{ICSEcounterfactual}  \cite{ICSEcounterfactual} used code counterfactuals, minimal changes to the source code under which the model ``changes its mind''. This approach helps developers understand which specific mutations influence the model's predictions. 

\vspace{-0.3cm}
\section{Methodology}
Our methodology is structured around three research questions. For RQ1, we aim to determine whether and how SPP to the code inputs influence the consistency of ACR tools' output. To do so, our research method involves: (I)~curating and leveraging a real-world dataset of Java code revision instances, (II)~designing and introducing SPP to Java methods to be revised, (III)~selecting five transformer-based ACR tools, and (IV)~designing an approach to evaluate the sensitivity of ACR tools against perturbations. For RQ2, we investigate which features of perturbations explain variations in model consistency. To do so, our research method involves (V)~extracting a set of features capturing structural and lexical properties of the perturbations, and (VI)~using a regression model to explain variations in model consistency. For RQ3, we explore (VII)~mitigation strategies aimed at improving model consistency. Each component is detailed in the following subsections.

\subsection{Dataset Preparation}

We base our evaluation on a dataset $\mathcal{D} = \{(c_i, r_i, \hat{c}_i)\}_{i=1}^{N}$ consisting of $N = 2032$ instances selected from the test split of a dataset consisting of method-level code review data collected from open-source GitHub projects, originally curated by~\citet{tufano2022}. Each instance consists of a triplet $(c_i, r_i, \hat{c}_i)$, where $c_i$ denotes the \textit{original code}, $r_i$ the \textit{review comment}, and $\hat{c}_i$ the corresponding \textit{human-written revised code}. The portion of $c_i$ targeted by $r_i$ is explicitly marked with \texttt{<START>} and \texttt{<END>} tags to indicate the span referenced by the comment. This dataset was also used in prior work to fine-tune T5-small and the LoRA-tuned LLaMA-7B models on its training split and assess their performances on its test split~\cite{tufano2022, lamaReviewer}. For our study, we select $N = 2032$ instances from this test split that are \textit{solvable} by at least one of the ACR tools included in our evaluation--for each instance, at least one ACR tool successfully generates $\hat{c}_i$ for ${c}_i$. This ensures that all selected instances represent code-review scenarios where automated code revision is feasible. Our evaluation is conducted at the method level. This choice is motivated by the fact that some of the ACR tools evaluated in this paper (e.g., T5-based and LoRA-tuned LLaMA models) are already fine-tuned on method-level inputs, and using the same input granularity ensures compatibility and a fair comparison across all tools. Since several of the evaluated ACR tools are pretrained on large-scale code corpora, we cannot rule out that an arbitrary input instance---or one of our perturbed variants---may coincidentally appear in their training data. However, we consider this risk low, because our perturbations systematically alter the surface form and token distribution of methods in ways that are unlikely to occur naturally in a pretraining corpus.

We define a set of $K = 9$ SPP functions $\mathcal{P} = \{p_1, p_2, \dots, p_K\}$, where each $p_k: (c_i, \hat{c}_i) \mapsto (c_i^{(k)}, \hat{c}_i^{(k)})$ produces perturbed code variant $c_i^{(k)}$ and $\hat{c}_i^{(k)}$, while preserving their semantic equivalence with $c_i$ and  $\hat{c}_i$ respectively (see Section \ref{perturbationstool}). Each perturbation modifies the structure or syntax of the methods while maintaining identical program behavior. 
For each applicable perturbation $p_k$, we generate a perturbed triplet $(c_i^{(k)}, r_i, \hat{c}_i^{(k)})$, yielding a perturbed dataset $\mathcal{D}' = \bigcup_{i=1}^{N} \{(c_i^{(k)}, r_i, \hat{c}_i^{(k)}) \mid p_k \text{ applicable to } c_i \}$. This results in over \num{10}K perturbed but semantically equivalent variants. The one-to-one mapping between a variant and its perturbation type allows attribution of model failures to specific perturbations.

Given a model $f_\theta$ that predicts a revised code $\tilde{c}_i = f_\theta(c_i, r_i)$, we define a successful revision (Exact Match; EXM) as when the model's generated revision $\tilde{c}_i$ exactly matches the human-written revision provided as the ground truth in the dataset (Exact Match; EXM): ${EXM}_i = 1$ when $\tilde{c}_i = \hat{c}_i$. The overall EXM accuracy for model $f_\theta$ is $\text{EXM}(f_\theta, \mathcal{D}) = \frac{1}{|\mathcal{D}|} \sum_{i=1}^{|\mathcal{D}|} \text{EXM}_i$, which ensures that all evaluated examples are solvable by at least one ACR tool under normal (unperturbed) conditions.

After generating perturbed variants for the $N = 2032$ review instances, we first evaluate each ACR tool individually to establish a reliable baseline of its performance on the original, unperturbed dataset $\mathcal{D}$. This step is essential because consistency can be meaningfully assessed when the model is demonstrably capable of solving the unperturbed task. For each model $f_\theta$, we define its \textit{solvable subset} as $\mathcal{S}_\theta = \{(c_i, r_i, \hat{c}_i) \in \mathcal{D} \mid f_\theta(c_i, r_i) = \hat{c}_i \}$. This per-model subset ensures that consistency is evaluated only on instances the model can successfully solve under unperturbed conditions. We then evaluate the same model on the \emph{perturbed variants} derived from subset  $\mathcal{S}_\theta' = \{(c_i^{(k)}, r_i, \hat{c}_i^{(k)}) \mid (c_i, r_i, \hat{c}_i) \in \mathcal{S}_\theta,\, p_k \text{ applicable} \}$, allowing us to measure performance degradation. Finally, to allow fair comparison across models, we constructed an \textit{intersection subset} $\mathcal{S}_{\cap} = \bigcap_{\theta} \mathcal{S}_\theta,$ comprising $|\mathcal{S}_{\cap}| = 464$ review instances that all models successfully revised on the original code. This shared subset provides a consistent reference point for cross-model comparison and removes baseline performance differences as a confounding factor. Consequently, any observed degradation in performance under perturbed inputs reflects true inconsistency rather than differences in model strength, architecture, or fine-tuning data.

\subsection{Semantics-preserving Perturbations (SPP)} \label{perturbationstool}

Using the Programming Concept, we designed a set of $K = 9$ types of SPP for Java methods, each aligned with one of three fundamental programming concepts: control flow, data flow, and identifier naming~\cite{controlFlow,dataFlow,Naming}. Programming Concept refers to three aspects: (i) how a method executes (control flow), (ii) how values are produced and propagated (data flow), and (iii) how identifiers are assigned and referenced (naming). These perturbations were iteratively developed through several rounds of examination and trial-and-error to ensure that each transformation is designed to preserve program semantics across a wide variety of Java methods. Following collaborative discussions among all authors, we manually inspected more than 100 representative Java methods and their perturbed variants across the nine SPP types, compiled transformed code, and verified behavioral equivalence to guarantee that the perturbations did not alter the program's logical behavior, input–output relations, or observable side effects. \autoref{fig:consistency_matters} shows an example through the try–catch wrapping perturbation. In this transformation, the body of a method is enclosed within a \texttt{try-catch} block whose \texttt{catch} clause rethrows any caught exception. Because the exception is rethrown, the program's termination remains identical to the original. The review comment still targets the same code region, and the intended fix does not change. This demonstrates how SPP can structurally modify code without affecting its meaning. For visualized examples of all perturbation types, please refer to the visual diffs provided in the \texttt{VisualPerturbs/} directory of our replication package~\cite{perturbationsReplication2025}. To systematically create these semantically equivalent variants of each instance $(c_i, r_i, \hat{c}_i) \in \mathcal{D}$, we developed an automated framework that applies controlled structural perturbations $p_k \in \mathcal{P}$ to Java code while preserving program semantics. The framework (i)~employs JavaParser (V3.23.1) to parse each $c_i$ into its Abstract Syntax Tree (AST), (ii)~modifies the AST to apply the selected applicable perturbations $p_k$, and (iii)~regenerates the perturbed source code $c_i^{(k)}$ from the modified AST. Because JavaParser enforces strict syntactic correctness, rejecting both ill-formed source code and invalid ASTs, this bidirectional parse–transform–serialize process guarantees that all perturbed outputs are syntactically valid and executable in Java. Each variant $c_i^{(k)}$ is produced by applying exactly one perturbation type, enabling precise attribution of any change in a model’s revision behavior to the corresponding perturbation $p_k$. By isolating the effect of individual perturbation, we can assess how different perturbation types influence model consistency without confounding interactions between multiple concurrent code changes. We structure semantically equivalent perturbations into 3 higher-level programming concepts as follows: 

\noindent\textbf{\controlflow.} Perturbations that involve modifying the structure of the control statements

    \noindent$\bullet$~\textit{If else swap} ($p_1$): Negates the conditions in the if statement and swaps the bodies of if-else blocks. If a method contains multiple \texttt{if–else} statements, we apply this transformation to all \texttt{if–else} blocks.

    \noindent$\bullet$~\textit{Dead Exception insertion} ($p_2$): Inserts a boolean variable `var' initialized to `false', and an if statement that throws an exception if `var' is true. Since var is always false, the if block is never executed.

    \noindent$\bullet$~\textit{Dead variable assignment insertion} ($p_3$): Inserts a boolean variable `var' initialized to `false', along with an if statement that reassigns it (if (var) { var = true; }).

    \noindent$\bullet$~\textit{Try and catch wrapper} ($p_4$): Wraps the entire method body in a try-catch block and rethrows any caught exceptions, preserving the original exception propagation behaviour. Because the catch clause immediately rethrows the same exception and no additional logic is executed, the method’s outputs and externally observable side effects remain unchanged. Methods already wrapped in a try-catch block are excluded.

    \noindent$\bullet$~\textit{Independent line swap} ($p_5$): Swaps any two adjacent statements (either variable declarations or assignments) only when the second statement does not rely on variables declared or assigned in the first statement or does not contain a method call.



\smallskip
\noindent\textbf{\dataflow.} Perturbations that involve altering how data is passed or stored.

    \noindent$\bullet$~ \textit{Return via variable} ($p_6$): Replace direct return statements with an intermediate variable declaration of the same return type, assigning the return value to that variable before returning it. This perturbation is applied only when the return type is not void, not Runnable, and not a generic wrapper of Void.

    \noindent$\bullet$~\textit{Def-use break} ($p_7$): Breaks the original definition-use chain of a variable by inserting a new variable---using the same initializer but a randomly generated 5-letter name---immediately after the original variable definition. All subsequent uses of the original variable are then replaced with this new variable name. If a method contains multiple variable definitions, we apply this transformation to all such variables and update their corresponding subsequent uses consistently within the method.

\smallskip
\noindent\textbf{\identifier.} Perturbations that change variable names.

In \identifier perturbations, changes to the code can affect variable names mentioned in the corresponding review comment. To preserve the integrity of the perturbation and ensure consistency between code and comment, we apply the same replacements to the review comment as well.

 \noindent$\bullet$~\textit{Random variable names} ($p_8$): Replaces variable names with randomly generated variable names (5 randomly selected letters), testing the impact of naming on code comprehension.  This perturbation requires at least one variable in the method to apply.

   \noindent$\bullet$~\textit{Shuffle variable names} ($p_9$): Randomly reassigns all local variables within the method to different names from a shuffled pool of variable names already used in the code, and updates all references consistently. This perturbation requires at least two variables to enable the shuffle.

\subsection{Inference Setup Across Models}

For LoRA-tuned LLaMA, LLaMA-3.3-70 B-Instruct-Turbo, ChatGPT-3.5-Turbo, and DeepSeek-V3, we employed a zero-shot prompting strategy by deliberately excluding persona-based instructions. This approach has been evaluated by \citet{pornprasit2024fine} alongside other prompting techniques, using the same dataset used in our study, showing that prompts without a persona were most effective. We have set the \textit{temperature} \footnote{Temperature, which ranges from 0 to 2, affects the randomness of the output: higher values produce more varied results, while lower ones yield more consistent answers.} to 0.2 for more focused and deterministic responses, as recommended for code refinement and revision tasks. The T5 model is treated differently, as prompting is not applicable. The model is an encoder-decoder transformer that was pretrained and fine-tuned specifically for the code revision task and was not instruction-tuned. Therefore, we use \emph{direct inference} for T5 using a structured format that was used during model fine-tuning. Since these evaluated ACR tools are inherently non-deterministic, we control for sampling variance to isolate the effect of semantics-preserving perturbations. Specifically, for each instance, we first verify that a tool can solve the original (unperturbed) task at least once under our fixed generation setting (among 10 generated revisions). We then evaluate the same tool on the corresponding semantically equivalent but structurally perturbed variants using the same setting, and measure whether it remains successful under these controlled structural changes.

\subsection{Evaluating Consistency}\label{sec:evaluation}
We distinguish between five representations of a review instance $(c_i, r_i, \hat{c}_i) \in \mathcal{D}$  used in our evaluation: (i)~\textit{original code} ($c_i$), the method as it appears in the dataset; (ii)~\textit{ground-truth revision} ($\hat{c}_i$),  the human-written revision implementing the reviewer's feedback; (iii)~\textit{perturbed code} ($c_i^{(k)}$),  a semantically equivalent variant of $c_i$  created through a controlled perturbation $p_k \in \mathcal{P}$;  (iv)~\textit{perturbed ground-truth revision} ($\hat{c}_i^{(k)}$), obtained by applying the same perturbation type $p_k$ applied to $c_i$ to create $c_i^{(k)}$; and (v)~\textit{model-generated revision for the perturbed input} ($\tilde{c}_i^{(k)}$), produced by the ACR tool when given $(c_i^{(k)}, r_i)$ as input.

The core goal of our evaluation is to measure the consistency of ACR tools by assessing whether a model can adapt its output to structural variations in the input code and still generate a correct revision. To ensure a fair comparison, the evaluation must account for the structural transformations introduced to the input code by the perturbations. When a model receives $(c_i^{(k)}, r_i)$ as input, it is expected to produce $\tilde{c}_i^{(k)}$ that mirrors the structural form of $c_i^{(k)}$. If the model instead generates $\hat{c}_i$, effectively ``undoing'' the perturbation, it indicates behavior dominated by memorized syntactic patterns rather than structural adaptability. If the model reverts $c_i^{(k)}$ back to the unperturbed structure, it is not only applying the requested edit, but implicitly normalizing the input into a form that resembles examples seen during training. This behavior suggests that the model's revision strategy may be tied to surface-level patterns rather than to the ability to localize the review intent and apply the corresponding change under structurally different, yet semantically equivalent, code contexts. To evaluate this behavior consistently, each perturbation function $p_k$ is designed to take both the original and revised code, applying the same SPP $p_k$  to both ${c}_i$ and $\hat{c}_i$, creating a structurally aligned perturbed reference $\hat{c}_i^{(k)}$. This step is necessary because the human revision $\hat{c}_i$ is written for the original method $c_i$, and therefore does not necessarily match the structure of the perturbed input $c_i^{(k)}$. By constructing $\hat{c}_i^{(k)}$, we ensure that the reference revision is structurally compatible with the perturbed input, so that the evaluation measures whether the model performs the intended edit under structural variation, rather than whether it reproduces a particular canonical formatting of the revision. This procedure preserves the integrity of the reference since all perturbations are semantics-preserving. Conceptually, this is equivalent to removing the perturbation from the model's generated revision before comparison, but we adopt the former for clarity and symmetry, ensuring that both the input and the reference share the same structural form when assessing model consistency. For example, in \autoref{fig:consistency_matters}, when the variant of the original code is wrapped inside a \texttt{try--catch} block, the ACR tool is expected to apply the reviewer's feedback within that same \texttt{try--catch} structure rather than reverting to the unwrapped form. If the reviewer-required change removes the syntactic opportunity for applying a given perturbation, or if the required change is itself equivalent to that perturbation, we exclude that instance from the dataset for the corresponding $p_k$ to avoid ambiguous comparisons. Considering the five representations of a review instance, we assess the consistency of ACR tools against SPP across four evaluation metrics as follows:

\paragraph{1. Exact Match (EXM)}
EXM measures whether the model-generated revision $\tilde{c}_i^{(k)}$  exactly matches the perturbed revision $\hat{c}_i^{(k)}$. It provides an upper bound on consistency, capturing cases where the model generates the expected structural and semantic changes. We use EXM mainly to measure consistency: If a model can revise the original code exactly as the human revision in the first place, with EXM we measure whether its output remains consistent over semantically equivalent perturbed versions. Also, EXM is the standard metric used in prior work~\cite{tufano2022, microsoft, nashaattowards, tufano2021towards, review4repair, coditt5, pornprasit2024fine, lu2023improving}
 and gives a strict lower bound on how often the model produces the intended human revision. Due to the strictness of EXM and its sensitivity to minor syntactic variations, we additionally report CodeBLEU and Edit Match to better capture partially correct revisions and structural similarity beyond exact string equality.
\paragraph{2. CodeBLEU}
We compute the CodeBLEU score~\cite{ren2020codebleu} between the model output $\tilde{c}_i^{(k)}$ and the perturbed ground-truth $\hat{c}_i^{(k)}$. A higher similarity with $\hat{c}_i^{(k)}$ indicates that the model successfully adapted to the structural variation in the input.

While EXM and CodeBLEU measure similarity between model-generated and ground-truth revisions, they may penalize ACR tools that correctly apply the intended fix but introduce additional unrelated changes. In such cases, even though the model performs the required revision correctly, textual differences cause Exact Match to fail and the CodeBLEU score to drop. To address this limitation, we introduce two complementary metrics. 

\paragraph{3. Edit Match (EM)} 
EM measures whether the model-generated revision $\tilde{c}_i^{(k)}$ includes all the edits made in the ground-truth revision $\hat{c}_i^{(k)}$. It is \textit{true} when $\tilde{c}_i^{(k)}$ contains every added and deleted code region present in $\hat{c}_i^{(k)}$, even if the model also introduces extra edits elsewhere. Thus, EM is satisfied when the model successfully performs all required changes, regardless of extra modifications. By definition, EM is always true when EXM is true, but EM can also hold when EXM fails. For example, in \autoref{fig:consistency_matters}, if the model correctly inserts the required null-check but also removes an unrelated \texttt{try–catch} block, EXM fails because of the extra deletion, but EM remains true since the essential revision, the addition of the null-check, is correctly applied.
Despite being more permissive than EXM, EM does not capture more idiomatic or efficient revisions. When a model produces a correct revision that requires fewer edits than the human ground truth, EM, same as EXM, will label it as incorrect.

\paragraph{4. Relative Edit Error (REE)} 
REE measures how much the model's edits deviate in magnitude from those in the ground-truth revision $\hat{c}_i^{(k)}$. It quantifies the proportional difference between the number of tokens modified when revising $c_i^{(k)} \!\rightarrow\! \tilde{c}_i^{(k)}$ and when revising $c_i^{(k)} \!\rightarrow\! \hat{c}_i^{(k)}$. In our evaluation, REE is computed only when EM is true. A value close to zero indicates that the model introduced approximately the same extent of modifications as the human developer; larger positive values indicate excessive or unnecessary edits. Formally, the Relative Edit Error is defined as:
\[
\text{REE} = \frac{|\text{edits}(c_i^{(k)} \!\rightarrow\! \tilde{c}_i^{(k)})| - |\text{edits}(c_i^{(k)} \!\rightarrow\! \hat{c}_i^{(k)})|}{|\text{edits}(c_i^{(k)} \!\rightarrow\! \hat{c}_i^{(k)})|}
\]

\subsection{Features of Perturbations}
\label{sec:features}

To further study the sensitivity of ACR tools to SPP, we extend our analysis to quantitative characteristics of the perturbations themselves. We hypothesize that ACR tools' behavior may depend not only on the categorical perturbation type $p_k$, but also on continuous or positional factors that affect input processing. Although semantically equivalent, perturbed snippets $c_i^{(k)}$  may differ from $c_i$ in properties such as length, position of the perturbation, or token distribution, factors that can affect the model's attention and context retention. For instance, perturbations near the lines targeted by a review comment $r_i$ may interfere with local attention, whereas those increasing code length may challenge context preservation. To capture these effects systematically, we define five measurable metrics describing distinct structural features of each perturbation. 

\noindent\textbf{Position (\textit{POS}).} Captures \textit{where} a perturbation occurs relative to the code span referenced by the review comment (the \textit{tagged span}, enclosed by \texttt{$\langle$START$\rangle$} and \texttt{$\langle$END$\rangle$} tags). Transformer-based models exhibit positional bias, under-attending to middle tokens~\cite{lostLengthPosition,positionBias,attentionMiddle,instructionposition}. A perturbation's position can thus affect model attention to the targeted span. We categorize each perturbation based on where its modified code (\textit{perturbed spans}) fall relative to the tagged span:  

{
\noindent\textit{Before:} All perturbed spans are before the tagged span. \\
\noindent\textit{After:} All perturbed spans are after the tagged span. \\
\textit{Inside:} All perturbed spans lie entirely within the tagged span. \\
\textit{Surrounding:} Perturbed spans exist both before and after the tagged span, without overlapping it. \\
\textit{Overlap-Before:} At least one perturbed span overlaps with the tagged span, and the rest are before it. \\
\textit{Overlap-After:} At least one perturbed span overlaps with the tagged span, and the rest are after it. \\
\textit{Overlap-Both:} Perturbed spans surround the tagged span, with at least one overlapping (i.e., the perturbation spans both sides and at least one part overlaps the tagged code).
}

\noindent\textbf{Perturbation Distance}. This feature measures the average of the shortest token-level distances between perturbed spans and the tagged span in the token space of $c_i^{(k)}$. We hypothesize that ACR tools are more sensitive to perturbations occurring near the reviewer-referenced code (tagged span). Using a custom static tokenizer that treats Java identifiers, keywords, operators, and the special tags \texttt{<START>} and \texttt{<END>} as tokens, we compute the shortest token-level distance between each perturbed span and the tagged span. Overlapping spans have a distance of zero. Each instance’s perturbation distance is calculated as the average of these shortest distances across all perturbed spans within $c_i^{(k)}$.

\smallskip
\noindent\textbf{\boldmath{$c_i \!\rightarrow\! c_i^{(k)}$ Token Edit.}}
This metric measures the number of token-level edits between the original code and its perturbed version, quantifying how much the input structure was changed. It counts the minimum number of token insertions, deletions, or substitutions required to transform $c_i$ into $c_i^{(k)}$. More edits indicate greater structural deviation, which we hypothesize increases the likelihood of model inconsistency despite unchanged semantics.

\smallskip
\noindent\textbf{\boldmath{$c_i^{(k)} \!\rightarrow\! \hat{c}_i^{(k)}$ Token Edit.}}
This metric measures the number of token-level edits between the perturbed code and its corresponding perturbed revision. It reflects the scale of change the model must apply to generate the correct revision. Smaller distances typically correspond to fewer and therefore easier edits needed to generate a revision based on model input; higher token edits represent more complex revision tasks that are more prone to failure.

\smallskip
\noindent\textbf{\boldmath{$c_i^{(k)}$ Length.}}
This feature records the total number of tokens in the perturbed input code. It captures the effect of input length on model behavior, as longer sequences increase processing load. Prior work~\cite{sameTaskLength,whyshort,ruler} shows that model accuracy declines as input length grows, making code length a relevant predictor of performance degradation.

\vspace{-0.1cm}
\subsection{Multi-level Mixed-effect logistic regression}

We model the probability that an ACR tool produces the exact match
revision after a perturbation with a \emph{multilevel mixed-effect logistic regression model} for a binary outcome
(\texttt{EXM} $\in\{0,1\}$) using a logit link.  The fixed part of the model includes five covariates that quantify distinct features of the perturbation (as defined in Section~\ref{sec:features}).  To account for baseline differences that are \emph{not} the focus of our inference, we add random intercepts for each (i) perturbation type and (ii) model used in each ACR tool. Formally, we define:

\vspace{-0.3cm}
{\small
\begin{equation}
\begin{aligned}
\text{logit}(\Pr[\text{Success}=1]) =\ & \beta_0 
+ \beta_1 \, \text{Pos} 
+ \beta_2 \, \text{PerturbationDist} \\
& + \beta_3 \, \text{TokenEdit}(c_i, c_i^{(k)}) \\
& + \beta_4 \, \text{TokenEdit}(c_i^{(k)}, \hat{c}_i^{(k)}) 
+ \beta_5 \, \text{Len}(c_i^{(k)}) \\
& + (1 \mid \text{PerturbationType}) \\
& + (1 \mid \text{LM})
\end{aligned}
\end{equation}}

Where $(1\mid\cdot)$ denotes a random intercept: each perturbation category and each LM receives its own baseline log-odds. The fixed‐effect coefficients~$\beta_1\!-\!\beta_5$ quantify the global effect of the perturbation features on consistency. To investigate the potential relationships among our predictor variables (i.e., the perturbation features), ensure model stability, and reduce the risk of multicollinearity, we adopted a three-step diagnostic procedure:

\noindent 
\noindent (i) We computed pairwise Spearman rank correlations among all continuous predictors (i.e., all features except for \texttt{Pos}, which is categorical). We considered $\rho > 0.7$ as an indicator of strong correlation that may warrant removal or transformation of a variable.
\noindent (ii) We calculated Variance Inflation Factors (VIF) for all continuous variables, using a threshold of 5 to detect problematic multicollinearity.
\noindent (iii)  For model fitting, we used the \texttt{glmer} function from the \texttt{lme4} package in R \cite{lme4}. We incrementally introduced the predictors into the model, monitoring coefficient stability to ensure that no variable disproportionately influenced the estimates of others. The results of these checks confirmed that none of the variables exhibited problematic multicollinearity. All Spearman correlation coefficients between predictor pairs remained well below the 0.7 threshold, and all VIF values were comfortably below the cutoff of 5. These diagnostics confirm that all selected covariates can be included in the final model without multicollinearity concerns, ensuring interpretability and robustness of the regression results.

\subsection{Mitigation: Changing Input Representation}\label{sec:prompting-strategies-methodology}
Based on prior findings on enhancing model reasoning and attention~\cite{pornprasit2024fine,wei2022chain}, we designed three mitigation strategies that change the input representation to emphasize and reinforce the span explicitly tagged for revision. These strategies aim to guide the model’s attention toward the reviewer-referenced code region and reduce its sensitivity to structural perturbations in other parts of the method.

\noindent$\bullet$~\textbf{Code Repetition (CR)}: Reinforces the revision target by repeating the code span between \texttt{<START>} and \texttt{<END>} inside the review comment. For all models, the review comment is rephrased using the format: \emph{``For this part of the Java code: \{tagged code span\}, this review comment is provided: \{original review comment\}.''}

\noindent$\bullet$~\textbf{Inline Commenting (IC)}: Embeds the review comment directly into the source code as an inline comment. The modified code includes the original review comment inserted immediately after the span marked between \texttt{<START>} and \texttt{<END>}. 

\noindent$\bullet$~\textbf{Chain-of-Thought (CoT)}: Encourages intermediate reasoning by appending the sentence \emph{``Provide step-by-step reasoning about how the review comment relates to the code''} . This strategy does not apply to T5, as T5 is not instruction-tuned; we therefore do not add CoT for T5.

\section{Results}

\subsection{\rqOneShort}

Since the dataset of \num{2032} review instances was constructed such that each instance is solvable by at least one of the evaluated ACR tools, and these ACR tools show varying results in terms of producing correct revision over this dataset, we assess their consistency at level of: (I) perturbed inputs created out of the subset of the original inputs for which the specific model has been able to produce an exact match ($\mathcal{S}_\theta$); (II) perturbed inputs created out of the subset of original input for which all the models have been able to produce an exact match ($\mathcal{S}_{\cap}$), which contains 464 out of the 2032 total inputs. Table~\ref{tab:exact_match_orig_stats} shows an overview of the number of inputs ($|\mathcal{S}_\theta|$) for which each model generated a revision that exactly matches the human-written ground-truth revision. To quantify the impact of perturbations, we calculate the relative drop in exact match from each model's baseline of 100\% performance ($\mathcal{S}_\theta$) across all types of perturbations. In other words, for a given model $m$, the maximum relative drop in consistency under perturbation is defined over the subset $\mathcal{S}_\theta$ as (where $\mathcal{P}$ is the set of perturbation types, and $\text{EXM}_p^{(m)}$ is the number of exact matches produced by model $m$ under perturbation $p$):

\begin{equation}
\max \Delta \text{EXM}^{(m)} (\%) = \max_{p \in \mathcal{P}} \left(1 - \text{EXM}_p^{(m)} \right) \times 100
\end{equation}

A similar drop $\max \Delta \text{EXM}{\cap}^{(m)}$ is computed over the shared subset $\mathcal{S}_{\cap}$, which includes 464 inputs for which all models originally succeeded. Lower values of $\Delta \text{EXM}^{(m)}$ indicate greater consistency of the ACR tool against the perturbations.

As shown in Table~\ref{tab:exact_match_orig_stats}, when evaluated on their respective baseline subsets ($\mathcal{S}_\theta$), the LoRA-tuned LLaMA model exhibits the largest performance degradation under semantics-preserving perturbations, with a maximum relative drop of 45.3\%. This indicates that, despite being fine-tuned specifically for code revision, it remains highly sensitive to structural variations in input code. DeepSeek-V3 follows with a substantial drop of 39.0\%.

\begin{table}[t]
    \caption{Exact match performance of ACR tools on the original input set ($\mathcal{D}$) and maximum observed performance drop under all types of SPP (\faLongArrowAltDown~$\max \Delta \text{EXM}$).
    }
    \centering
    \begin{tabular}{lrrrr}
     \toprule
        \textbf{Model}  & $|\mathcal{S}_\theta|$ & (\%) & \multicolumn{2}{c}{\faLongArrowAltDown~$\max \Delta \text{EXM}$~(\%)} \\
    \cmidrule(lr){4-5}
   & & & $\mathcal{S}_{\cap}$ & $\mathcal{S}_\theta$ \\
    \midrule
    T5                 & 1951 & 96.01 & 33.3 &  34.6\\
    LoRA-tuned LLaMA   & 1048 & 51.57 & 24.5 &  45.3\\
    LLaMA 3.3-70B      & 1190 & 58.56 & 24.5 &  27.7\\
    GPT-3.5 Turbo      & 1042 & 51.28 & 40.9 &  33.1\\
    DeepSeek V3        & 1103 & 54.28 & 33.3 &  39.0\\
    \bottomrule
    \end{tabular}
    \label{tab:exact_match_orig_stats}
\end{table}

In Table~\ref{tab:intersect_metrics_summary}, we report the performance of all ACR tools on the intersection subset ($\mathcal{S}_{\cap}$)—the 464 instances where every model originally achieved an exact match on the unperturbed input. The table presents the drop in average Exact Match ($\Delta$EXM), the drop in average Edit Match ($\Delta$EM), the average Relative Edit Error (REE), and the average CodeBLEU score for each perturbation type. Lower $\Delta$EXM and $\Delta$EM indicate higher consistency, while lower REE values show that even when a model fails to exactly reproduce the ground-truth revision, its additional edits were minimal. Both tuned models (fine-tuned T5 and LoRA-tuned LLaMA) show the greatest sensitivity to the control-flow perturbation type of $p_1$ based on both $\Delta$EXM and $\Delta$EM. The next largest degradation for LoRA-tuned LLaMA appears under data-flow perturbation type of $p_6$, indicating data-flow changes also disrupt its consistency. In contrast, both models remain moderately robust to dead-code insertions ($p_2$, $p_3$), where $\Delta$EM values are the smallest—showing that semantically inert changes have limited effect on edit alignment. However, the REE values show the opposite pattern and higher REE values suggest that while models rarely fail edit matching (EM), their few errors involve mild over-editing rather than complete misalignment when handling dead-code insertions. Larger foundation models—LLaMA 3.3-70B, GPT-3.5 Turbo, and DeepSeek-V3—exhibit very low REE across perturbations, suggesting less over-editing. However, these models suffer their highest $\Delta$EM and $\Delta$EXM under structurally complex transformations such as $p_1$ and $p_7$, confirming that changes in control and data dependencies remain the most challenging for all ACR tools. Since perturbations that target different types of programming fundamentals led, on average, to relatively similar impact, we deepen our analysis to uncover the common features between these perturbation types, which are responsible for what to expect in terms of consistency level.

\begin{table*}[t]
\centering
\caption{Performance of ACR tools on the intersection subset ($\mathcal{S}_{\cap}$) across perturbation types ($p_i$). Each column reports Exact Match ($\Delta$EXM), Edit Match ($\Delta$EM), Relative Edit Error (REE), and CodeBLEU.}
\vspace{-0.3cm}
\label{tab:intersect_metrics_summary}
\begin{adjustbox}{max width=\textwidth}
\rowcolors{1}{white}{blue!4}
\begin{tabular}{l|rrrr|rrrr|rrrr|rrrr|rrrr}
\toprule
\multirow{2}{*}{\textbf{$\mathcal{P}$}} & \multicolumn{4}{c|}{\textbf{T5}} & \multicolumn{4}{c|}{\textbf{LoRA-tuned LLaMA}} & \multicolumn{4}{c|}{\textbf{LLaMA 3.3-70B}} & \multicolumn{4}{c|}{\textbf{GPT-3.5 Turbo}} & \multicolumn{4}{c}{\textbf{DeepSeek V3}} \\
\cmidrule(lr){2-21}
& \faLongArrowAltDown $\Delta$EXM & \faLongArrowAltDown $\Delta$EM & REE & CBLEU & \faLongArrowAltDown $\Delta$EXM  & \faLongArrowAltDown $\Delta$EM & REE & CBLEU & \faLongArrowAltDown $\Delta$EXM  & \faLongArrowAltDown $\Delta$EM  & REE & CBLEU & \faLongArrowAltDown $\Delta$EXM  & \faLongArrowAltDown $\Delta$EM  & REE & CBLEU & \faLongArrowAltDown $\Delta$EXM & \faLongArrowAltDown $\Delta$EM & REE & CBLEU \\
\midrule
$p_1$ & 33.3 & 16.1 & 0.260 & 0.680 & 45.2 & 14.04 & 0.519 & 0.580 & 20.4 & 15.1 & 0.107 & 0.740 & 40.9 & 25.8 & 0.160 & 0.560 & 30.1 & 21.5 & 0.260 & 0.650 \\
$p_2$ & 14.7 & 4.5 & 0.410 & 0.880 & 21.2 & 4.5 & 0.594 & 0.807 & 7.5 & 6.0 & 0.018 & 0.920 & 9.7 & 7.2 & 0.080 & 0.890 & 22.7 & 15.5 & 0.300 & 0.750 \\
$p_3$ & 22.2 & 3.7 & 0.570 & 0.810 & 21.2 & 4.5 & 0.728 & 0.680 & 6.5 & 5.2 & 0.024 & 0.920 & 12.0 & 7.5 & 0.090 & 0.850 & 23.7 & 16.0 & 0.280 & 0.710 \\
$p_4$ & 11.0 & 3.8 & 0.190 & 0.900 & 22.7 & 3.3 & 0.248 & 0.806 & 6.0 & 6.0 & 0.030 & 0.910 & 11.8 & 9.0 & 0.060 & 0.860 & 15.6 & 14.2 & 0.130 & 0.780 \\
$p_5$ & 11.1 & 11.1 & 0.000 & 0.910 & 22.2 & 11.1 & 0.062 & 0.814 & 11.1 & 11.1 & 0.000 & 0.810 & 22.2 & 11.1 & 0.050 & 0.840 & 33.3 & 11.1 & 0.250 & 0.740 \\
$p_6$ & 33.3 & 4.0 & 0.610 & 0.690 & 31.3 & 4.0 & 0.598 & 0.727 & 11.1 & 2.0 & 0.090 & 0.880 & 20.2 & 8.1 & 0.250 & 0.780 & 19.2 & 16.2 & 0.110 & 0.700 \\
$p_7$ & 33.1 & 12.3 & 0.390 & 0.740 & 24.5 & 4.3 & 0.477 & 0.754 & 24.5 & 17.2 & 0.026 & 0.760 & 27.6 & 19.0 & 0.200 & 0.730 & 38.0 & 23.9 & 0.260 & 0.660 \\
$p_8$ & 16.8 & 10.1 & 0.270 & 0.840 & 8.4 & 3.5 & 0.115 & 0.91 & 8.1 & 6.9 & 0.040 & 0.880 & 15.0 & 10.1 & 0.090 & 0.820 & 11.6 & 11.0 & 0.110 & 0.800 \\
$p_9$ & 13.4 & 8.8 & 0.110 & 0.880 & 15.3 & 4.6 & 0.310 & 0.852 & 13.9 & 10.6 & 0.040 & 0.850 & 19.9 & 16.2 & 0.100 & 0.800 & 17.1 & 12.0 & 0.360 & 0.770 \\
\bottomrule
\end{tabular}

\end{adjustbox}
\end{table*}

\vspace{-0.2cm}
\subsection{\rqTwo}\label{logitresults}

We begin by examining the descriptive statistics of the perturbation features in Table~\ref{tab:perturbation_stats_combined}, which provide an overview of the variation in token edits, input lengths, and the distribution of positional categories across all perturbations. To assess which features of perturbations influence the likelihood that an ACR tool preserves an EXM, we conducted a multi-level mixed-effects logistic regression with random intercepts for perturbation type and model. The dependent variable is binary, indicating whether an exact match was preserved between model-generated revision $\tilde{c}_i^{(k)}$ and perturbed ground-truth revision $\hat{c}_i^{(k)}$. Because the EXM metric represents the strictest measure of consistency, it can be interpreted as the upper bound of consistency. Moreover, EXM exhibited trends in degradation that were consistent with those observed for other evaluation metrics (e.g., EM and CodeBLEU), but in a more conservative manner. Therefore, we selected EXM as the dependent variable. The logistic regression analysis results in Table~\ref{tab:fixed_effects_sorted} present predictors sorted by their effect size, quantified as the odds ratio (OR). The odds ratio represents the multiplicative change in the odds of preserving an exact match for a one-unit increase in the predictor variable, holding other variables constant. Values below 1 indicate a decrease in the likelihood of consistency (i.e., higher sensitivity to perturbations), whereas values above 1 indicate an increased likelihood of maintaining consistency. Among the predictors, perturbation distance exhibited the strongest effect size and was the only factor with an odds ratio (OR) greater than 1, indicating a positive and statistically significant influence on likelihood of EXM. This suggests that the further a perturbation occurs from the region of code targeted by the review comment (tagged span), the higher the likelihood that the ACR tool generates a revision exactly matching $\hat{c}_i^{(k)}$. In contrast, all other predictors had negative effects (OR < 1), indicating a reduced probability of consistency as their values increased. The most negative predictor was observed for perturbations located ``inside'' the tagged span, implying that when a perturbation directly affects the region of code targeted by the review comment, the likelihood of EXM decreases substantially across all models. Furthermore, while the marginal $R^2$ (0.059) indicates that the fixed effects explain a small portion of the variance (only 5.9\%), the conditional $R^2$ (0.220) means the model explains 22\% of the variation in the data after including the random effects for ACR tool and perturbation type. This shows that these random effects help the model better capture differences that come from using different ACR tools or types of perturbations, aspects that would not be explained by the fixed effects alone.

\begin{table*}[t]
\centering
\small
\caption{Descriptive statistics of numerical features and position distribution of perturbations across all perturbation types}
\label{tab:perturbation_stats_combined}
\begin{adjustbox}{max width=\textwidth}
\begin{tabular}{lrrrrr@{\hskip 1.5cm}lr@{\hskip 0.5cm}lr}
\toprule
\multicolumn{6}{c}{\textbf{Descriptive Statistics of Numerical features}} & \multicolumn{4}{c}{\textbf{Position Category Distribution}} \\
\cmidrule(lr){1-6} \cmidrule(lr){7-10}
\textbf{Metric} & \textbf{Min} & \textbf{Max} & \textbf{Average} & \textbf{Median} & \textbf{Std Dev} 
& \textbf{Position Category} & \textbf{Cases} & \textbf{Position Category} & \textbf{Cases} \\
\midrule
Perturbation Distance & 0.00 & 217.00 & 15.42 & 8.50 & 20.42
& Before           & 3418 & Overlap-Before   & 453 \\
$c_i \!\rightarrow\! c_i^{(k)}$ Token Edit & 1.00 & 180.00 & 9.82 & 9.00 & 8.33
& Surrounding      & 2710 & Overlap-After    & 394 \\
$c_i^{(k)} \!\rightarrow\! \hat{c}_i^{(k)}$ Token Edit & 1.00 & 166.00 & 5.86 & 3.00 & 9.37
& Overlap-Both     & 1864 & Inside           & 137 \\
Perturbed Input Length & 7.00 & 341.00 & 58.95 & 46.00 & 42.03 
& After            & 1389 &                  &      \\
\bottomrule
\end{tabular}
\end{adjustbox}
\end{table*}

\begin{table}[t]
\centering
\caption{Mixed-effects logistic regression results for predicting exact match consistency under perturbations}
\label{tab:fixed_effects_sorted}
\begin{adjustbox}{width=\columnwidth}
\rowcolors{1}{white}{blue!4}
\begin{tabular}{lrrrrr}
\toprule
\textbf{Predictor} & \textbf{Estim.} & \textbf{OR} & \textbf{95\% CI} & \textbf{Pr($>|z|$)} \\
\midrule
(Intercept)              &  1.359 & 3.894 & (2.704 , 5.608) & 2.80e-13\textsuperscript{***} \\
Perturbation Distance    &  0.118 & 1.125 & (1.076 , 1.176) & 1.85e-07\textsuperscript{***} \\
$c_i \!\rightarrow\! c_i^{(k)}$ Token Edit & -0.181 & 0.835 & (0.809 , 0.862) & \textless{}2e-16\textsuperscript{***} \\
$POS$ (After)            & -0.197 & 0.821 & (0.747 , 0.902) & 3.99e-05\textsuperscript{***} \\
$POS$ (Overlap After)    & -0.229 & 0.795 & (0.674 , 0.939) & 6.72e-03\textsuperscript{**} \\
$c_i^{(k)} \!\rightarrow\! \hat{c}_i^{(k)}$ Token Edit & -0.335 & 0.715 & (0.694 , 0.736) & \textless{}2e-16\textsuperscript{***} \\
$POS$ (Surrounding)      & -0.342 & 0.710 & (0.644 , 0.784) & 1.01e-11\textsuperscript{***} \\
$POS$ (Overlap Before)   & -0.565 & 0.568 & (0.493 , 0.656) & 1.12e-14\textsuperscript{***} \\
$POS$ (Overlap Both)     & -0.568 & 0.566 & (0.506 , 0.634) & \textless{}2e-16\textsuperscript{***} \\
$POS$ (Inside)           & -0.690 & 0.502 & (0.399 , 0.632) & 4.46e-09\textsuperscript{***} \\
Perturbed Input Length   & -0.019 & 0.981 & (0.937 , 1.027) & 0.41906 \\
\midrule
\end{tabular}
\end{adjustbox}

\begin{adjustbox}{width=\columnwidth}
\begin{tabular}{lrlr}
\textbf{Marginal $R^2$} & 0.059 & \textbf{Conditional $R^2$} & 0.220 \\
\textbf{\# of observations} & \num{33122} & \textbf{Groups} & 9 PT, 5 ACR tools \\
\bottomrule
\end{tabular}
\end{adjustbox}

\vspace{0.1cm}
\begin{adjustbox}{max width=\columnwidth}
\begin{tabular}{ll}
Significance & \textsuperscript{***}\,$p < 0.001$,
\textsuperscript{**}\,$p < 0.01$,
\textsuperscript{*}\,$p < 0.05$ \\
PT & Perturbation Types
\end{tabular}
\end{adjustbox}
\end{table}

\begin{table*}[t]
\centering
\caption{Impact of Mitigation Strategies on ACR Tools for each perturbation type ($p_i$)}
\vspace{-0.2cm}
\smallskip
\label{tab:mitigation_impact}
\begin{adjustbox}{max width=\textwidth}
\begin{tabular}{l
*{5}{!{\vrule width 0.4pt}rrrr}}
\toprule
& \multicolumn{4}{c!{\vrule width 0.4pt}}{\textbf{T5}} 
& \multicolumn{4}{c!{\vrule width 0.4pt}}{\textbf{LoRA-tuned LLaMA}} 
& \multicolumn{4}{c!{\vrule width 0.4pt}}{\textbf{LLaMA 3.3-70B}} 
& \multicolumn{4}{c!{\vrule width 0.4pt}}{\textbf{GPT-3.5 Turbo}} 
& \multicolumn{4}{c}{\textbf{DeepSeek V3}} \\
\textbf{$\mathcal{P}$}
& Default & $\Delta$CoT & $\Delta$CR & $\Delta$IC
& Default & $\Delta$CoT & $\Delta$CR & $\Delta$IC
& Default & $\Delta$CoT & $\Delta$CR & $\Delta$IC
& Default & $\Delta$CoT & $\Delta$CR & $\Delta$IC
& Default & $\Delta$CoT & $\Delta$CR & $\Delta$IC  \\
\midrule
$p_1$ & 66.7 & -- & -10.8 & -34.4 & 54.8 & \cellcolor{lightgreen}+2.2 & -3.2 & -12.9 & 79.6 & \cellcolor{lightgreen}+7.1 & -1.1 & -16.1 & 59.1 & -9.7 & \cellcolor{lightgreen}+1.1 & -24.7 & 69.9 & -0.4 & 0.0 & -10.8 \\
$p_2$ & 85.3 & -- & -18.0 & -49.9 & 78.8 & \cellcolor{lightgreen}+1.0 & -6.5 & -5.7 & 92.5 & \cellcolor{lightgreen}+3.0 & 0.0 & -15.0 & 90.3 & -19.0 & \cellcolor{lightgreen}+1.5 & -37.2 & 77.3 & -7.8 & \cellcolor{lightgreen}+7.2 & -2.5 \\

$p_3$ & 77.8 & -- & -22.9 & -46.4 & 68.1 & 0.0 & -3.7 & \cellcolor{lightgreen}+0.5 & 93.5 & \cellcolor{lightgreen}+2.3 & 0.2 & -16.5 & 88.0 & -19.0 & -0.7 & -34.9 & 76.3 & -8.1 & \cellcolor{lightgreen}+7.0 & -5.5 \\

$p_4$ & 89.0 & -- & -20.5 & -40.5 & 77.3 & \cellcolor{lightgreen}+0.3 & -5.8 & -1.4 & 94.0 & -2.4 & -1.1 & -13.2 & 88.2 & -7.4 & \cellcolor{lightgreen}+0.3 & -28.2 & 84.4 & -3.8 & \cellcolor{lightgreen}+2.5 & -5.2 \\

$p_5$ & 88.9 & -- & -22.2 & -55.6 & 77.8 & 0.0 & 0.0 & 0.0 & 88.9 & \cellcolor{lightgreen}+11.1 & -11.1 & -33.3 & 77.8 & 0.0 & 0.0 & -33.3 & 66.7 & \cellcolor{lightgreen}+13.3 & \cellcolor{lightgreen}+11.1 & 0.0 \\

$p_6$ & 66.7 & -- & -13.1 & -35.4 & 68.7 & \cellcolor{lightgreen}+10.1 & -7.1 & \cellcolor{lightgreen}+8.1 & 88.9 & \cellcolor{lightgreen}+8.0 & \cellcolor{lightgreen}+5.1 & -2.0 & 79.8 & -3.0 & \cellcolor{lightgreen}+8.1 & -20.2 & 80.8 & -11.9 & \cellcolor{lightgreen}+1.0 & -5.1 \\

$p_7$ & 66.9 & -- & -9.2 & -31.3 & 71.8 & -2.5 & -9.8 & -9.2 & 75.5 & -2.8 & -1.8 & -12.3 & 72.4 & -9.8 & -1.8 & -24.5 & 62.0 & -6.9 & \cellcolor{lightgreen}+3.1 & -2.5 \\

$p_8$ & 83.2 & -- & -16.2 & -38.7 & 91.6 & -1.2 & -6.1 & -9.8 & 91.9 & \cellcolor{lightgreen}+1.2 & -4.3 & -15.3 & 85.0 & -16.2 & -1.2 & -28.0 & 88.4 & -4.6 & -1.7 & -4.6 \\

$p_9$ & 86.6 & -- & -13.4 & -39.8 & 84.7 & -1.9 & -6.5 & -1.4 & 86.1 & -0.9 & -1.9 & -12.5 & 80.1 & -16.2 & 0.0 & -27.8 & 82.9 & -5.6 & -0.9 & -1.9 \\

\midrule
\rowcolor{gray!20}
\textbf{Min} & 89.0 & -- & -9.2 & -31.3 & 91.6 & 10.1 & 0.0 & 8.1 & 94.0 & 11.1 & 5.1 & -2.0 & 90.3 & 0.0 & 8.1 & -20.2 & 88.4 & 13.3 & 11.1 & 0.0 \\

\rowcolor{gray!10}
\textbf{Max} & 66.7 & -- & -22.9 & -55.6 & 54.8 & -2.5 & -9.8 & -12.9 & 75.5 & -2.8 & -11.1 & -33.3 & 59.1 & -19.0 & -1.8 & -37.2 & 62.0 & -11.9 & -1.7 & -10.8 \\

\rowcolor{gray!5}
\textbf{Avg} & 79.0 & -- & -16.3 & -41.3 & 74.8 & 0.9 & -5.4 & -3.5 & 87.9 & 3.0 & -1.8 & -15.1 & 80.1 & -11.1 & 0.8 & -28.8 & 76.5 & -4.0 & 3.3 & -4.2 \\

\bottomrule
\end{tabular}
\end{adjustbox}
\end{table*}

\vspace{-0.2cm}
\subsection{\rqThree}

Table \ref{tab:mitigation_impact} reports the percentage of average EXM for each model in the default (no-mitigation) condition on baseline subsets ($\mathcal{S}_\theta$), followed by the $\Delta$ values representing the change in EXM percentage under each mitigation strategy—Chain-of-Thought (CoT), Code Repetition (CR), and Inline Comment (IC). The results are shown for each perturbation type $\mathcal{P}$ and each model \(m\), allowing a detailed assessment of how each strategy affects model consistency under specific perturbations. The values are computed over the subset $\mathcal{S}_\theta$. Since our goal is not to compare models against each other but rather to assess the impact of mitigation techniques, this within-model evaluation is sufficient. It allows for isolating the effect of mitigation strategies on model consistency, independent of differences in baseline performance. In general, the improvements, highlighted as green positive values, are modest across models, indicating that input-representation changes have a limited impact on reducing inconsistency. However, a pattern emerges across architectures: LLaMA-based models tend to show small but consistent gains when Chain-of-Thought prompting is applied. In contrast, GPT-3.5 Turbo and DeepSeek V3 exhibit their most consistent, but minor, improvements under the Code Repetition (CR) strategy, where the lines of code referenced by the review comment are repeated within the comment itself. While these few isolated cases show slight improvements, the overall trend across perturbation types and models reveals that all three input-representation strategies generally reduce consistency. As shown in Table~\ref{tab:mitigation_impact}, the majority of  $\mathcal{S}_\theta$ values for Chain-of-Thought, Code Repetition, and Inline Comment are negative, indicating that these strategies more often decrease rather than enhance exact-match performance. This suggests that introducing additional textual or structural information into the input—whether through reasoning prompts, code repetition, or inline annotations—can inadvertently distract the model's attention, leading to less stable behavior when confronted with semantically equivalent code variants. In other words, while intended to strengthen the linkage between review feedback and the relevant code region, these interventions amplify inconsistencies by increasing complexity or shifting token-level alignment between review comments and code.

\section{Discussion}

Our study evaluates the consistency of transformer-based ACR tools when applied to semantics-preserving perturbations in Java source code. We highlight aspects relevant to developers aiming to integrate ACR tools into practical workflows as well as to researchers investigating the creation and evolution of these tools.

\noindent\textbf{\noindent \textit{On the variability in revision consistency.}} The analysis reveals significant variability in how different ACR tools maintain consistency under SPP. For instance, we observed substantial changes reaching up to 45.3\% for LoRA-tuned LLaMA across all perturbations on its solvable subset, and up to 40.9\% for GPT-3.5 Turbo on the intersection subset. Beyond exact match (EXM), our Edit Match (EM) metric reveals that tuned models (T5 and LoRA-tuned LLaMA) exhibit the greatest sensitivity to control-flow perturbations such as $p_1$, with EM drops of 16.1\% and 14.04\% respectively, while remaining moderately robust to dead-code insertions ($p_2$, $p_3$). The Relative Edit Error (REE) metric further shows that larger foundation models (LLaMA 3.3-70B, GPT-3.5 Turbo, and DeepSeek V3) exhibit very low REE across perturbations, indicating less over-editing when they do generate revisions, whereas tuned models show higher REE values (up to 0.728 for LoRA-tuned LLaMA on $p_3$), suggesting more over-editing on semantically inert changes. Also, high-performing models, such as DeepSeek V3 and GPT-3.5 Turbo, do not provide consistent outputs. This result may stem from LLMs' reliance on surface-level token patterns rather than semantic understanding.  As autoregressive models, LLMs often retrieve answers from training correlations~\cite{brown2020language,zhang2022opt}, making them susceptible to minimal context shifts~\cite{moradi2021evaluating,zhu2023promptrobust}. These findings align with prior observations on LLM instability under semantically neutral input changes for other types of tasks~\cite{zhu2023promptrobust,CounterfactualsForCode,lin2022truthfulqa}.

\noindent\textbf{\textit{On the factors influencing consistency. }}
Our multilevel mixed-effects logistic regression identifies key factors significantly impacting ACR tool sensitivity. Perturbations occurring within the focal revision span (region of code targeted by the review comment) have the strongest impact, meaning that when a perturbation directly affects the code regions targeted by the review comment, the odds of maintaining an exact match decrease by approximately 50\%. Odds of maintaining an exact match in perturbations involving higher token edits (i.e., cases requiring greater syntactic divergence between the perturbed source and its correct revision) drop by 28.5\%. Conversely, perturbations farther from the focal revision span are less disruptive. These results show the impact of \textit{the proximity and magnitude of changes} relative to the focal revision span under review. Because LLMs are highly sensitive to input structure, even semantically neutral modifications within the focal revision span may modify the model's internal representations, affecting its ability to maintain a consistent fix.

\noindent \textbf{\textit{On the effectiveness of input representations.}} We investigated improving consistency by strategically changing models' input representation, including code repetition, and inline comments, as well as Chain-of-Thought (CoT) prompting. Contrary to our expectations, these techniques did not enhance consistency. Inline comments, for instance, generally resulted in a significant decrease in consistency (up to a 55.6\% drop) for T5 on $p_5$, followed by (up to a 37.2\% drop) for GPT-3.5 Turbo on $p_2$ and (up to a 33.3\% drop) for LLaMa 3.3-70B on $p_5$. Code repetition occasionally improved consistency in the case of GPT-3.5 Turbo and DeepSeek V3. We conclude for the obtained results that adoption of these mitigation techniques seem to degrade the consistency of models rather then improving it, and underline the necessity of a different approach. Especially in code-related tasks, such noise can interfere with token alignment, attention calibration, or decoding behavior learned~\cite{chen2021evaluating}.

\noindent \textbf{\textit{Recommendations for future work.}} 
Moving beyond standard performance metrics and integrating perturbation-based evaluations in model validation processes can help better understand the consistency of the measured results. Developers should test their tools rigorously against perturbations occurring near or within the focal revision segment, as our analysis revealed these perturbations have the highest negative impact on model predictions. Prompting strategies commonly used in NLP tasks are not directly transferable to code revision and can even reduce consistency; thus, avoid one-size-fits-all prompting and prompt design should be context-aware and rigorously tailored to the hierarchical and positional nature of code. Moreover, future research could explore architectural or training-level interventions towards more detailed, contextually adaptive mitigation approaches tailored specifically to code review scenarios. Finally, exposing consistency indicators or robustness scores to end users can foster more informed adoption of ACR tools in real-world code review workflows. Ultimately, achieving reliable and interpretable ACR systems will require advancing from models that mimic code patterns to those that reason over semantics and context with higher consistency.

\section{Threats to Validity}
\noindent\textbf{Internal Validity.} 
Our evaluation assumes that each SPP changes only the surface structure of a method while preserving its runtime behavior, input–output relations, and externally observable side effects. However, semantic preservation cannot be guaranteed universally: in few edge cases, a transformation that is intended to be semantics-preserving may interact with the reviewed issue itself or with corner-case program behavior. We use four complementary metrics (EXM, EM, REE, CodeBLEU) to measure consistency. While EXM is strict, EM and REE address cases where models apply correct fixes but introduce additional changes. We control for correctness by evaluating only inputs where original predictions matched the ground truth. Our evaluation constructs a perturbed ground-truth revision $\hat{c}_i^{(k)}$ by applying the same perturbation $p_k$ to both the original code $c_i$ and the human revision $\hat{c}_i$. This implicitly assumes that a developer would implement an equivalent revision when the input is presented in a different yet semantically equivalent structural form. In practice, developers may reasonably produce alternative revisions that are correct but structurally different from $\hat{c}_i^{(k)}$, especially if the perturbation changes code readability or highlights different refactoring opportunities. As a result, some model outputs may be penalized despite being valid. We partially mitigate this risk by reporting edit-based metrics in addition to Exact Match, but our evaluation prioritizes consistency with the human reference revision rather than capturing the full space of functionally correct revisions.

\noindent\textbf{External Validity.} 
Several evaluated ACR tools are pretrained on large-scale code corpora, and we cannot rule out the possibility that a code review instance in our evaluation may appear in their training data. Such exposure could inflate absolute performance. Our findings are based on Java methods extracted from real-world GitHub code reviews and a set of five representative transformer-based ACR tools. While our perturbation operators are designed to capture general programming concepts (control flow, data flow, and identifier naming) and could be adapted to other languages, the generalizability of our results may be limited by the specific dataset characteristics, programming language, and tool selection considered in this study.

\section{Conclusion}
LLMs are increasingly being integrated into software development processes, including critical tasks such as code review and code revision. As their role becomes more central in the development pipeline, we aim to understand the reliability and consistency of their output. In the study presented in this paper, we investigated the consistency of LLM-based ACR tools by evaluating their sensitivity to SPP (i.e., syntactic and structural changes that maintain the original program behavior). Our goal was to assess the extent to which these models produce consistent responses when the meaning of the input code remains unchanged---a key property for dependable automation. By designing nine types of SPPs, we evaluated the consistency of five ACR tools on over 10K perturbed variants derived from real-world GitHub code reviews. Our results reveal that model consistency is far from guaranteed: the ability of ACR tools to provide consistent output drops by up to 45.3\%. Sensitivity is most pronounced when perturbations occur inside or near the code region referenced by the review comment, thus providing additional evidence that model attention is heavily localized and surface-dependent. 
We also evaluated mitigation strategies that modify the input representation (including code repetition, inline comments, and Chain-of-Thought prompting) to guide model attention toward the review target and to test their potential in stabilizing model responses. However, these strategies produced limited and inconsistent benefits, with occasional gains for specific model families (e.g., LLaMA family with Chain-of-Thought, GPT-3.5 Turbo and DeepSeek V3 with code repetition) but no overall improvement. Our findings highlight a critical gap between strong performance measured on benchmarks and robustness in slightingly modified, yet realistic scenarios.
Our study provides further evidence that perturbation-based evaluation frameworks can be a valuable complement to accuracy metrics, by enabling developers and researchers to diagnose model brittleness.

\subsection*{Acknowledgements}
 
S. Pirouzkhah, S. Serbout, and A. Bacchelli gratefully acknowledge the support of the Swiss National Science Foundation through the SNF Projects No. 200021\_197227 and No. 200021M\_205146. 
 
\clearpage

\bibliographystyle{ACM-Reference-Format}
\bibliography{reference}

@inproceedings{tufano2022,
  title={Using pre-trained models to boost code review automation},
  author={Tufano, Rosalia and Masiero, Simone and Mastropaolo, Antonio and Pascarella, Luca and Poshyvanyk, Denys and Bavota, Gabriele},
  booktitle={Proceedings of the 44th International Conference on Software Engineering},
  pages={2291--2302},
  year={2022}
}

@article{raffel2020exploring,
  title={Exploring the limits of transfer learning with a unified text-to-text transformer},
  author={Raffel, Colin and Shazeer, Noam and Roberts, Adam and Lee, Katherine and Narang, Sharan and Matena, Michael and Zhou, Yanqi and Li, Wei and Liu, Peter J},
  journal={Journal of machine learning research},
  volume={21},
  number={140},
  pages={1--67},
  year={2020}
}

@article{vaswani2017attention,
  title={Attention is all you need},
  author={Vaswani, Ashish and Shazeer, Noam and Parmar, Niki and Uszkoreit, Jakob and Jones, Llion and Gomez, Aidan N and Kaiser, {\L}ukasz and Polosukhin, Illia},
  journal={Advances in neural information processing systems},
  volume={30},
  year={2017}
}

@article{hochreiter1997long,
  title={Long short-term memory},
  author={Hochreiter, Sepp and Schmidhuber, J{\"u}rgen},
  journal={Neural computation},
  volume={9},
  number={8},
  pages={1735--1780},
  year={1997},
  publisher={MIT press}
}

@inproceedings{vincentatscale,
  title={Towards automating code review at scale},
  author={Hellendoorn, Vincent J and Tsay, Jason and Mukherjee, Manisha and Hirzel, Martin},
  booktitle={Proceedings of the 29th ACM Joint Meeting on European Software Engineering Conference and Symposium on the Foundations of Software Engineering},
  pages={1479--1482},
  year={2021}
}

@inproceedings{tufano2021towards,
  title={Towards automating code review activities},
  author={Tufano, Rosalia and Pascarella, Luca and Tufano, Michele and Poshyvanyk, Denys and Bavota, Gabriele},
  booktitle={2021 IEEE/ACM 43rd International Conference on Software Engineering (ICSE)},
  pages={163--174},
  year={2021},
  organization={IEEE}
}

@inproceedings{microsoft,
  title={Automating code review activities by large-scale pre-training},
  author={Li, Zhiyu and Lu, Shuai and Guo, Daya and Duan, Nan and Jannu, Shailesh and Jenks, Grant and Majumder, Deep and Green, Jared and Svyatkovskiy, Alexey and Fu, Shengyu and others},
  booktitle={Proceedings of the 30th ACM Joint European Software Engineering Conference and Symposium on the Foundations of Software Engineering},
  pages={1035--1047},
  year={2022}
}

@inproceedings{Howfar,
  title={Generation-based code review automation: How far are we?},
  author={Zhou, Xin and Kim, Kisub and Xu, Bowen and Han, DongGyun and He, Junda and Lo, David},
  booktitle={2023 IEEE/ACM 31st International Conference on Program Comprehension (ICPC)},
  pages={215--226},
  year={2023},
  organization={IEEE}
}

@article{brown2020language,
  title={Language models are few-shot learners},
  author={Brown, Tom and Mann, Benjamin and Ryder, Nick and Subbiah, Melanie and Kaplan, Jared D and Dhariwal, Prafulla and Neelakantan, Arvind and Shyam, Pranav and Sastry, Girish and Askell, Amanda and others},
  journal={Advances in neural information processing systems},
  volume={33},
  pages={1877--1901},
  year={2020}
}

@article{zhang2022opt,
  title={Opt: Open pre-trained transformer language models},
  author={Zhang, Susan and Roller, Stephen and Goyal, Naman and Artetxe, Mikel and Chen, Moya and Chen, Shuohui and Dewan, Christopher and Diab, Mona and Li, Xian and Lin, Xi Victoria and others},
  journal={arXiv preprint arXiv:2205.01068},
  year={2022}
}

@inproceedings{moradi2021evaluating,
  title={Evaluating the Robustness of Neural Language Models to Input Perturbations},
  author={Moradi, Milad and Samwald, Matthias},
  booktitle={Proceedings of the 2021 Conference on Empirical Methods in Natural Language Processing},
  pages={1558--1570},
  year={2021}
}

@inproceedings{lin2022truthfulqa,
  title={TruthfulQA: Measuring How Models Mimic Human Falsehoods},
  author={Lin, Stephanie and Hilton, Jacob and Evans, Owain},
  booktitle={Proceedings of the 60th Annual Meeting of the Association for Computational Linguistics (Volume 1: Long Papers)},
  pages={3214--3252},
  year={2022}
}

@article{chen2021evaluating,
  title={Evaluating large language models trained on code},
  author={Chen, Mark and Tworek, Jerry and Jun, Heewoo and Yuan, Qiming and Pinto, Henrique Ponde De Oliveira and Kaplan, Jared and Edwards, Harri and Burda, Yuri and Joseph, Nicholas and Brockman, Greg and others},
  journal={arXiv preprint arXiv:2107.03374},
  year={2021}
}

@inproceedings{zhu2023promptrobust,
  title={Promptrobust: Towards evaluating the robustness of large language models on adversarial prompts},
  author={Zhu, Kaijie and Wang, Jindong and Zhou, Jiaheng and Wang, Zichen and Chen, Hao and Wang, Yidong and Yang, Linyi and Ye, Wei and Zhang, Yue and Gong, Neil and others},
  booktitle={Proceedings of the 1st ACM Workshop on Large AI Systems and Models with Privacy and Safety Analysis},
  pages={57--68},
  year={2023}
}

@misc{Patanamon,
  title={AutoTransform: Automated Code Transformation to Support Modern Code Review Process. In 2022 IEEE/ACM 44st International Conference on Software Engineering (ICSE)},
  author={Thongtanunam, Patanamon and Pornprasit, Chanathip and Tantithamthavorn, Chakkrit},
  year={2022},
  publisher={IEEE}
}

@inproceedings{edman2024cute,
  title={CUTE: Measuring LLMs’ Understanding of Their Tokens},
  author={Edman, Lukas and Schmid, Helmut and Fraser, Alexander},
  booktitle={Proceedings of the 2024 Conference on Empirical Methods in Natural Language Processing},
  pages={3017--3026},
  year={2024}
}

@inproceedings{bender2021dangers,
  title={On the dangers of stochastic parrots: Can language models be too big?},
  author={Bender, Emily M and Gebru, Timnit and McMillan-Major, Angelina and Shmitchell, Shmargaret},
  booktitle={Proceedings of the 2021 ACM conference on fairness, accountability, and transparency},
  pages={610--623},
  year={2021}
}

@inproceedings{bacchelliexpectations,
  title={Expectations, outcomes, and challenges of modern code review},
  author={Bacchelli, Alberto and Bird, Christian},
  booktitle={2013 35th International Conference on Software Engineering (ICSE)},
  pages={712--721},
  year={2013},
  organization={IEEE}
}

@inproceedings{CounterfactualsForCode,
  title={Do large code models understand programming concepts? counterfactual analysis for code predicates},
  author={Hooda, Ashish and Christodorescu, Mihai and Allamanis, Miltiadis and Wilson, Aaron and Fawaz, Kassem and Jha, Somesh},
  booktitle={Forty-first International Conference on Machine Learning},
  year={2024}
}

@inproceedings{ICSEcounterfactual,
  title={Counterfactual explanations for models of code},
  author={Cito, J{\"u}rgen and Dillig, Isil and Murali, Vijayaraghavan and Chandra, Satish},
  booktitle={Proceedings of the 44th international conference on software engineering: software engineering in practice},
  pages={125--134},
  year={2022}
}

@article{controlFlow,
  title={Control flow analysis},
  author={Allen, Frances E},
  journal={ACM Sigplan Notices},
  volume={5},
  number={7},
  pages={1--19},
  year={1970},
  publisher={ACM New York, NY, USA}
}

@article{dataFlow,
  title={Data flow analysis in software reliability},
  author={Fosdick, Lloyd D and Osterweil, Leon J},
  journal={ACM Computing Surveys (CSUR)},
  volume={8},
  number={3},
  pages={305--330},
  year={1976},
  publisher={ACM New York, NY, USA}
}

@inproceedings{Naming,
  title={Evaluation of software understandability based on fuzzy matrix},
  author={Lin, Jin-Cherng and Wu, Kuo-Chiang},
  booktitle={2008 IEEE International Conference on Fuzzy Systems (IEEE World Congress on Computational Intelligence)},
  pages={887--892},
  year={2008},
  organization={IEEE}
}

@inproceedings{sadowski2018modern,
  title={Modern code review: a case study at google},
  author={Sadowski, Caitlin and S{\"o}derberg, Emma and Church, Luke and Sipko, Michal and Bacchelli, Alberto},
  booktitle={Proceedings of the 40th international conference on software engineering: Software engineering in practice},
  pages={181--190},
  year={2018}
}

@article{lostLengthPosition,
  title={Lost-in-the-Middle in Long-Text Generation: Synthetic Dataset, Evaluation Framework, and Mitigation},
  author={Zhang, Junhao and Zhang, Richong and Kong, Fanshuang and Miao, Ziyang and Ye, Yanhan and Zheng, Yaowei},
  journal={arXiv preprint arXiv:2503.06868},
  year={2025}
}

@article{sameTaskLength,
  title={Same task, more tokens: the impact of input length on the reasoning performance of large language models},
  author={Levy, Mosh and Jacoby, Alon and Goldberg, Yoav},
  journal={arXiv preprint arXiv:2402.14848},
  year={2024}
}

@article{whyshort,
  title={Why Does the Effective Context Length of LLMs Fall Short?},
  author={An, Chenxin and Zhang, Jun and Zhong, Ming and Li, Lei and Gong, Shansan and Luo, Yao and Xu, Jingjing and Kong, Lingpeng},
  journal={arXiv preprint arXiv:2410.18745},
  year={2024}
}

@article{ruler,
  title={RULER: What's the Real Context Size of Your Long-Context Language Models?},
  author={Hsieh, Cheng-Ping and Sun, Simeng and Kriman, Samuel and Acharya, Shantanu and Rekesh, Dima and Jia, Fei and Zhang, Yang and Ginsburg, Boris},
  journal={arXiv preprint arXiv:2404.06654},
  year={2024}
}

@article{positionBias,
  title={Position-aware parameter efficient fine-tuning approach for reducing positional bias in llms},
  author={Zhang, Zheng and Yang, Fan and Jiang, Ziyan and Chen, Zheng and Zhao, Zhengyang and Ma, Chengyuan and Zhao, Liang and Liu, Yang},
  journal={arXiv preprint arXiv:2404.01430},
  year={2024}
}

@article{instructionposition,
  title={Instruction position matters in sequence generation with large language models},
  author={Liu, Yijin and Zeng, Xianfeng and Meng, Fandong and Zhou, Jie},
  journal={arXiv preprint arXiv:2308.12097},
  year={2023}
}

@article{attentionMiddle,
  title={Attention instruction: Amplifying attention in the middle via prompting},
  author={Zhang, Meiru and Meng, Zaiqiao and Collier, Nigel},
  journal={arXiv preprint arXiv:2406.17095},
  year={2024}
}

@inproceedings{zhang2024attacks,
  title={Attacks and Defenses for Large Language Models on Coding Tasks},
  author={Zhang, Chi and Wang, Zifan and Zhao, Ruoshi and Mangal, Ravi and Fredrikson, Matt and Jia, Limin and Pasareanu, Corina},
  booktitle={Proceedings of the 39th IEEE/ACM International Conference on Automated Software Engineering},
  pages={2268--2272},
  year={2024}
}

@article{wang2022recode,
  title={ReCode: Robustness evaluation of code generation models},
  author={Wang, Shiqi and Li, Zheng and Qian, Haifeng and Yang, Chenghao and Wang, Zijian and Shang, Mingyue and Kumar, Varun and Tan, Samson and Ray, Baishakhi and Bhatia, Parminder and others},
  journal={arXiv preprint arXiv:2212.10264},
  year={2022}
}

@inproceedings{lamaReviewer,
  title={Llama-reviewer: Advancing code review automation with large language models through parameter-efficient fine-tuning},
  author={Lu, Junyi and Yu, Lei and Li, Xiaojia and Yang, Li and Zuo, Chun},
  booktitle={2023 IEEE 34th International Symposium on Software Reliability Engineering (ISSRE)},
  pages={647--658},
  year={2023},
  organization={IEEE}
}

@article{SLRCR,
  title={Automating Code Review: A Systematic Literature Review},
  author={Tufano, Rosalia and Bavota, Gabriele},
  journal={arXiv preprint arXiv:2503.09510},
  year={2025}
}

@article{distilledT,
  title={Improving the learning of code review successive tasks with cross-task knowledge distillation},
  author={Ben Sghaier, Oussama and Sahraoui, Houari},
  journal={Proceedings of the ACM on Software Engineering},
  volume={1},
  number={FSE},
  pages={1086--1106},
  year={2024},
  publisher={ACM New York, NY, USA}
}

@article{nashaattowards,
  title={Towards efficient fine-tuning of language models with organizational data for automated software review},
  author={Nashaat, Mona and Miller, James},
  journal={IEEE Transactions on Software Engineering},
  year={2024},
  publisher={IEEE}
}

@article{review4repair,
  title={Review4repair: Code review aided automatic program repairing},
  author={Huq, Faria and Hasan, Masum and Haque, Md Mahim Anjum and Mahbub, Sazan and Iqbal, Anindya and Ahmed, Toufique},
  journal={Information and Software Technology},
  volume={143},
  pages={106765},
  year={2022},
  publisher={Elsevier}
}

@inproceedings{coditt5,
  title={Coditt5: Pretraining for source code and natural language editing},
  author={Zhang, Jiyang and Panthaplackel, Sheena and Nie, Pengyu and Li, Junyi Jessy and Gligoric, Milos},
  booktitle={Proceedings of the 37th IEEE/ACM International Conference on Automated Software Engineering},
  pages={1--12},
  year={2022}
}

@article{pornprasit2024fine,
  title={Fine-tuning and prompt engineering for large language models-based code review automation},
  author={Pornprasit, Chanathip and Tantithamthavorn, Chakkrit},
  journal={Information and Software Technology},
  volume={175},
  pages={107523},
  year={2024},
  publisher={Elsevier}
}

@inproceedings{lu2023improving,
  title={Improving Code Refinement for Code Review Via Input Reconstruction and Ensemble Learning},
  author={Lu, Jiawei and Tang, Zhijie and Liu, Zhongxin},
  booktitle={2023 30th Asia-Pacific Software Engineering Conference (APSEC)},
  pages={161--170},
  year={2023},
  organization={IEEE}
}

@inproceedings{frommgen2024resolving,
  title={Resolving code review comments with machine learning},
  author={Fr{\"o}mmgen, Alexander and Austin, Jacob and Choy, Peter and Ghelani, Nimesh and Kharatyan, Lera and Surita, Gabriela and Khrapko, Elena and Lamblin, Pascal and Manzagol, Pierre-Antoine and Revaj, Marcus and others},
  booktitle={Proceedings of the 46th International Conference on Software Engineering: Software Engineering in Practice},
  pages={204--215},
  year={2024}
}

@misc{lme4,
  title={lme4: Mixed-effects modeling with R},
  author={Bates, Douglas M},
  year={2010},
  publisher={Springer}
}

@article{wei2022chain,
  title={Chain-of-thought prompting elicits reasoning in large language models},
  author={Wei, Jason and Wang, Xuezhi and Schuurmans, Dale and Bosma, Maarten and Xia, Fei and Chi, Ed and Le, Quoc V and Zhou, Denny and others},
  journal={Advances in neural information processing systems},
  volume={35},
  pages={24824--24837},
  year={2022}
}

@misc{perturbationsReplication2025,
  author       = {S Pirouzkhah},
  title        = {Replication Package},
  year         = {2025},
  url          = {https://github.com/shirinpirouzkhaah/PerturbationsEvaluation},
  note         = {}
}

@inproceedings{moralesquality,
  title={Do code review practices impact design quality? a case study of the qt, vtk, and itk projects},
  author={Morales, Rodrigo and McIntosh, Shane and Khomh, Foutse},
  booktitle={2015 IEEE 22nd international conference on software analysis, evolution, and reengineering (SANER)},
  pages={171--180},
  year={2015},
  organization={IEEE}
}

@inproceedings{bavotaquality,
  title={Four eyes are better than two: On the impact of code reviews on software quality},
  author={Bavota, Gabriele and Russo, Barbara},
  booktitle={2015 IEEE International Conference on Software Maintenance and Evolution (ICSME)},
  pages={81--90},
  year={2015},
  organization={IEEE}
}

@article{improvequality,
  title={Software inspections: an effective verification process},
  author={Ackerman, A. Frank and Buchwald, Lynne S. and Lewski, Frank H.},
  journal={IEEE software},
  volume={6},
  number={3},
  pages={31--36},
  year={1989},
  publisher={IEEE}
}

@inproceedings{rigbydefect,
  title={Open source software peer review practices: a case study of the apache server},
  author={Rigby, Peter C and German, Daniel M and Storey, Margaret-Anne},
  booktitle={Proceedings of the 30th international conference on Software engineering},
  pages={541--550},
  year={2008}
}

@inproceedings{reducedefects,
  title={Software inspections and the industrial production of software},
  author={Ackerman, A Frank and Fowler, Priscilla J and Ebenau, Robert G},
  booktitle={Proc. of a symposium on Software validation: inspection-testing-verification-alternatives},
  pages={13--40},
  year={1984}
}

@article{ren2020codebleu,
  title={Codebleu: a method for automatic evaluation of code synthesis},
  author={Ren, Shuo and Guo, Daya and Lu, Shuai and Zhou, Long and Liu, Shujie and Tang, Duyu and Sundaresan, Neel and Zhou, Ming and Blanco, Ambrosio and Ma, Shuai},
  journal={arXiv preprint arXiv:2009.10297},
  year={2020}
}

@article{touvron2023llama,
  title={Llama: Open and efficient foundation language models},
  author={Touvron, Hugo and Lavril, Thibaut and Izacard, Gautier and Martinet, Xavier and Lachaux, Marie-Anne and Lacroix, Timoth{\'e}e and Rozi{\`e}re, Baptiste and Goyal, Naman and Hambro, Eric and Azhar, Faisal and others},
  journal={arXiv preprint arXiv:2302.13971},
  year={2023}
}

@article{hu2022lora,
  title={Lora: Low-rank adaptation of large language models.},
  author={Hu, Edward J and Shen, Yelong and Wallis, Phillip and Allen-Zhu, Zeyuan and Li, Yuanzhi and Wang, Shean and Wang, Lu and Chen, Weizhu and others},
  journal={ICLR},
  volume={1},
  number={2},
  pages={3},
  year={2022}
}

@article{dubey2024llama,
  title={The llama 3 herd of models},
  author={Dubey, Abhimanyu and Jauhri, Abhinav and Pandey, Abhinav and Kadian, Abhishek and Al-Dahle, Ahmad and Letman, Aiesha and Mathur, Akhil and Schelten, Alan and Yang, Amy and Fan, Angela and others},
  journal={arXiv e-prints},
  pages={arXiv--2407},
  year={2024}
}

@misc{openai2023gpt35,
  title={GPT-3.5 Turbo},
  author={OpenAI},
  year={2023},
  url={https://platform.openai.com/docs/models/gpt-3-5-turbo},
  note={Accessed: 2024-01-15}
}

@article{deepseekai2024deepseekv3,
  title={DeepSeek-V3 Technical Report},
  author={{DeepSeek-AI}},
  journal={arXiv preprint arXiv:2412.19437},
  year={2024}
}
\end{document}